\documentclass[journal]{IEEEtran}
\usepackage{cite}

\ifCLASSINFOpdf
  \usepackage[pdftex]{graphicx}
\else
\fi
\begin{document}
%
\title{Electrically driven photonic crystal nanocavity devices}
%
%
%

\author{Gary~Shambat, Bryan~Ellis, Jan~Petykiewicz, Marie~A.~Mayer, Arka~Majumdar, Tomas~Sarmiento,
        James~Harris, Eugene~E.~Haller, and Jelena Vu\v{c}kovi\'{c}
\thanks{G.~Shambat, B.~Ellis, J.~Petykiewicz, Arka~Majumdar, Tomas~Sarmiento, James~Harris, and J.~Vu\v{c}kovi\'{c} are with the Department of Electrical Engineering, Stanford University, Stanford, CA 94305 USA (e-mail: gshambat@stanford.edu; bryane@stanford.edu; janp@stanford.edu; arkam@stanford.edu; tsarmie@stanford.edu; harris@snowboard.stanford.edu; jela@stanford.edu).}
\thanks{M.~A.~Mayer and E.~E.~Haller are with the Materials Sciences Division, Lawrence Berkeley National Laboratory, Berkeley, CA 94720 USA (e-mail: mamayer@berkeley.edu; eehaller@lbl.gov}}

%
%

\markboth{}%
{Shell \MakeLowercase{\textit{et al.}}: Bare Demo of IEEEtran.cls for Journals}
%



\maketitle

\begin{abstract}
Interest in photonic crystal nanocavities is fueled by advances in device performance, particularly in the development of low-threshold laser sources. Effective electrical control of high performance photonic crystal lasers has thus far remained elusive due to the complexities associated with current injection into cavities. A fabrication procedure for electrically pumping photonic crystal membrane devices using a lateral p-i-n junction has been developed and is described in this work. We have demonstrated electrically pumped lasing in our junctions with a threshold of 181 nA at 50K - the lowest threshold ever demonstrated in an electrically pumped laser. At room temperature we find that our devices behave as single-mode light-emitting diodes (LEDs), which when directly modulated, have an ultrafast electrical response up to 10 GHz corresponding to less than 1 fJ/bit energy operation - the lowest for any optical transmitter. In addition, we have demonstrated electrical pumping of photonic crystal nanobeam LEDs, and have built fiber taper coupled electro-optic modulators. Fiber-coupled photodetectors based on two-photon absorption are also demonstrated as well as multiply integrated components that can be independently electrically controlled. The presented electrical injection platform is a major step forward in providing practical low power and integrable devices for on-chip photonics. 

\end{abstract}


%
\IEEEpeerreviewmaketitle

\section{Introduction}
%
%
%
%
\IEEEPARstart{P}{hotonic} crystal (PC) nanocavities have been the focus of intense research in recent years as these engineered nanostructures have opened the door for novel physics and device applications. A tremendous amount of progress has been made both in optimizing cavity properties such as the quality (Q) factor \cite{jelena1, akahane1, noda1} and mode volume as well as in developing interesting applications that are intrinsically enabled through a nanoscale dielectric form factor \cite{noda2, painter1, jelena2}. Early work in this field showed that localized defect modes can be created by perturbing the periodicity of a photonic lattice, creating highly customizable nanocavities with simple control over mode field patterns, radiation profiles, spectral positioning, and photon lifetime \cite{fan1, painter2}. The first experimental demonstration of a PC cavity laser began a wave of research in active photonic crystal cavity devices \cite{painter3}. PC nanocavity lasers have advanced remarkably and now represent the state of the art in low-threshold lasers \cite{noda3}. In such high Q-factor and small mode-volume cavities the Purcell factor can be quite high, reducing the lasing threshold and increasing the modulation rate \cite{yamamoto}. Optically pumped PC nanocavity lasers have been demonstrated to have thresholds of only a few nW, high output powers, and modulation rates exceeding 100 GHz \cite{strauf, matsuo, altug}. Furthermore, they can operate in continuous wave mode at room temperature and can be efficiently coupled to passive waveguides for optoelectronic integrated circuit applications \cite{nomura, nozaki}.  

Although sophisticated PC lasers and active devices have been developed, their corresponding electrical control has lagged tremendously. Because of the challenges associated with electrical pumping of photonic crystal membranes, all of the aforementioned laser demonstrations relied on impractical optical pumping.  There has been one previous demonstration of an electrically pumped photonic crystal laser using a vertical p-i-n junction grown within the semiconductor membrane \cite{park1}. A current post is used to inject carriers into the cavity region; however the current post limits the quality factor of the cavity, restricts the choice of the cavity design, and requires a complicated fabrication procedure \cite{park2}. In addition, a high threshold current of 260 $\mu$A was observed, significantly higher than in optically pumped PC devices and even exceeding that of good vertical-cavity surface-emitting lasers (VCSELs) \cite{macdougal}. Other groups have used similar vertical p-i-n designs for electrical pumping of PC LEDs but have not been able to resolve the complications associated with this inefficient injection platform \cite{francardi, chakravarty}. Due to the limited current spreading ability of the thin conductive layers, most of the electroluminescence (EL) is not coupled to the cavity, wasting electrical power and heating the device.  

We have therefore devised a new method for electrical control of photonic crystal cavities employing a lateral p-i-n junction (shown schematically in Fig. 1(a) and visualized in the SEM in Fig. 1(b)) \cite{ellisled, ellislaser}. The geometry inherent to 2D photonic crystal membranes lends itself more easily to lateral current injection defined by ion implanted p- and n-type regions. This device architecture has a simple and flexible fabrication procedure with high control over the current flow. In addition, arbitrary PC cavity designs with high Q-factors can be used along with coupling waveguides for efficient light extraction. Moreover, doping is introduced only in desired areas, enabling efficient integration of active and passive devices (as opposed to vertical p-i-n junctions where the doped regions are defined during growth, and in the entire wafer). While the most immediate application of our lateral junction is for nanocavity lasers and light-emitting diodes, the architecture is suitable for any number of applications needing active, integrable control of photonic crystal cavities. Finally, such a platform offers an a new degree of freedom for electrical control or tuning beyond that demonstrated in prior devices manipulated by electrostatics or local fields \cite{loncar1, andrei1}. 

Our paper is organized as follows. Section II goes over the device fabrication details and simulated electrical behavior. In section III, we present our electrically driven nanocavity laser results at cryogenic temperatures. Section IV discusses room temperature operation and ultrafast modulation of our single-mode LED. Section V describes additional applications of our lateral junction in nanobeam photonic crystal LEDs and fiber taper-coupled devices such as electro-optic modulators, photodetectors, and multiply integrated components. Finally, section VI summarizes our results and provides future directions for work in this area. 

\begin{figure}[!t]
\centering
\includegraphics[width=3.5in]{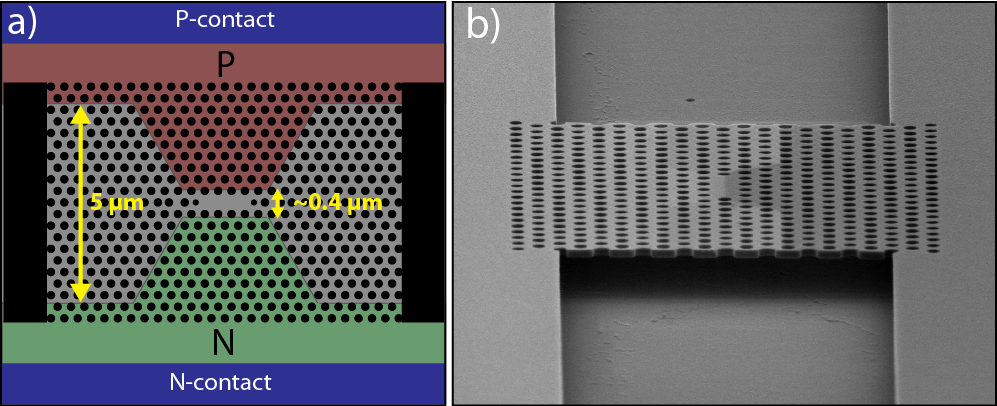}
\caption{(a) Schematic of the electrically driven photonic crystal cavity devices. The p-type doping region is indicated in red, and the n-type region in green. The width of the intrinsic region is narrow in the center to direct current flow to the cavity. The doping profile is tapered to ensure proper electrical injection. A trench is added to the sides of the cavity to reduce leakage current. (b) Tilted SEM of a fabricated laterally doped structure. A faint outline of the doping regions is visible.}
\label{fig_sim}
\end{figure}

\section{Fabrication and Design}

As mentioned above, our solution to the PC cavity electrical injection problem is to use a lateral junction formed by ion implantation \cite{ellisled}. Lateral current injection (LCI) has become routine for silicon-based electro-optic ring modulators in recent years owing to mature device process knowledge \cite{lipson}. More recently, a lateral junction in Si photonic crystals has been demonstrated for compact modulators and detectors \cite{tanabe1, tanabe2}. On the other hand, far less attention has been put towards III-V processed lateral junctions. It is possible to form lateral junctions in III-V materials during molecular beam epitaxy growth by incorporating appropriate dopant ions during multiple regrowth steps. However, this technique is very complicated and requires additional electron-beam lithography steps \cite{watanabe}. Alternatively, ion implantation has been explored as a method to form LCI edge-emitting GaAs lasers and more recently electroluminescent devices in InGaAsP PC membranes \cite{tager, long}. We build off of these works and refine a technique to form a lateral junction in GaAs with high precision and reproducibility.

We first note the choice of III-V membrane and active gain material, here gallium arsenide and indium arsenide self-assembled quantum dots (QDs). Quantum dots are prefered for the gain medium because they can have exceptionally low transparency carrier densities, allowing for reduced threshold lasing \cite{nomura}. The low transparency density also allows for relaxed constraints on the carrier injection levels and corresponding doping levels. Furthermore, nonradiative recombination of the dots themselves is improved versus similar quantum well (QW) systems \cite{englund}. 

Figure 2 shows a simplified schematic diagram of the lateral junction photonic crystal fabrication procedure. First, alignment marks were defined on the unpatterned wafer using electron beam lithography and a thick layer of silicon nitride was deposited on the sample to serve as a mask for ion implantation of Si. Electron beam lithography was used to pattern the n-type doping region and Si ions were implanted at an energy and dose such that the maximum of the doping density was  6 $\times$ 10$^{17}$ cm$^{-3}$ and the maximum of the dopant distribution was near the middle of the membrane. Another nitride mask was used to define the p-type doping region, formed by Be ions and achieving a doping density of 2.5 $\times$ 10$^{19}$ cm$^{-3}$. Next, a  tensile strained silicon nitride cap was deposited to prevent As out-diffusion during the subsequent high temperature anneal. The samples were then annealed in a rapid thermal annealer to activate the dopants and remove almost all of the lattice damage caused by the ion implantation. The photonic crystal pattern was defined using electron beam lithography and etched into the membrane. Simultaneous with the photonic crystal, trenches were etched to the sides of the cavity and all the way around each of the contacts; this was found to reduce the leakage current to reasonable levels. Finally, metal contacts were deposited in a lift-off process, activated, and the photonic crystal membranes were undercut. For full fabrication details see \cite{ellislaser}. 

\begin{figure}[!t]
\centering
\includegraphics[width=3.5in]{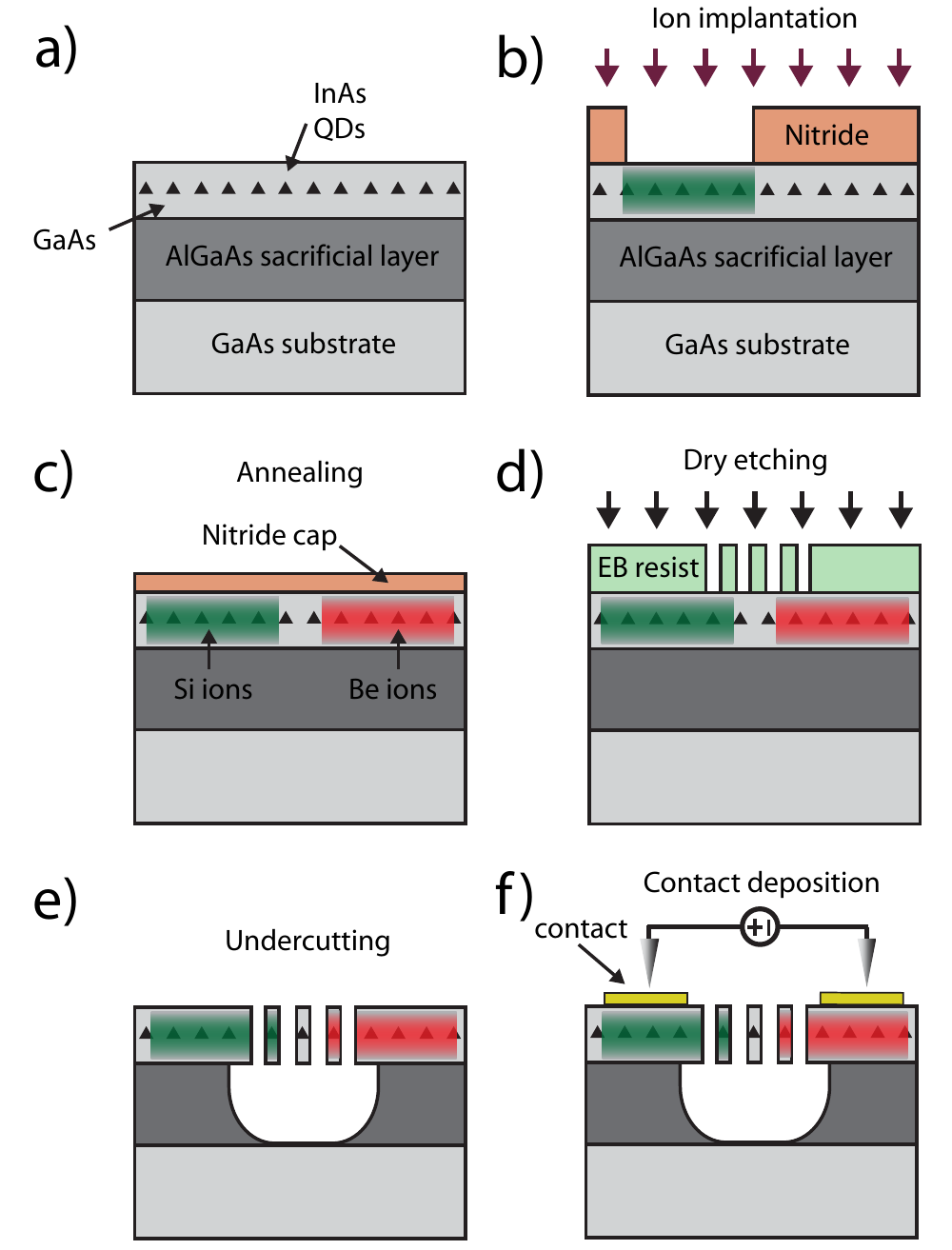}
\caption{Schematic illustration of the fabrication process (a) The starting material is a 220 nm thick membrane of GaAs with embedded InAs QDs. (b) Si and Be ions are implanted through a silicon nitride mask patterned by electron beam lithography. (c) The implanted dopants are activated by annealing at 850 $^{\circ}$C for 30s with a nitride cap which is subsequently removed by dry etching. (d) The photonic crystal pattern is defined in a resist by electron beam lithography and transferred into the GaAs membrane by dry etching. (e) The AlGaAs sacrificial layer is removed with wet etching. (f) The p and n contacts are deposited by photolithography and liftoff.}
\label{fig_sim}
\end{figure}

Si and Be ions were chosen because they offer the best combination of low lattice damage and high activation efficiency. In fact, our p- and n-type region activation carrier concentrations of 2.5 $\times$ 10$^{19}$ cm$^{-3}$ and 6 $\times$ 10$^{17}$ cm$^{-3}$ are close to the highest expected activation values in the literature for GaAs \cite{rao}. In our former study, we used Mg ions for the p region; however, we found activation carrier concentrations were lower at 3 $\times$ 10$^{18}$ cm$^{-3}$ \cite{ellisled}. Both the physical layout and activation concentration of our devices were found through scanning capacitance microscopy (SCM) \cite{williams}, as seen in Figure 3. The image shows that we have excellent control over the doping layout with an accuracy of within 30 nm determined by our e-beam alignment procedure. By comparing the position of the nitride mask edge to the measured doping edge, we can determine the distance of dopant diffusion and therefore control the intrinsic cavity region width. This parameter is very important for several reasons. Spatial overlap of the cavity mode profile with the heavily doped regions leads to free carrier absorption (FCA) and degradation of the Q-factor. In terms of realizing the best intrinsic Q-factor, the p and n regions would ideally be spaced out as far as possible. Counteracting this force is the fact that carrier injection into the intrinsic region goes down drastically as the doping regions are spaced farther apart, eventually to a value below the inversion condition for the QDs. This happens because the diffusion length for carriers in GaAs with a high non-radiative recombination rate is extremely short (200 nm for electrons and 40 nm for holes) \cite{shambatnanobeam}. Therefore a compromise between FCA and carrier injection must be made for a set intrinsic region width. Lastly, ion implantation degrades much of the QD emission even after the lattice is annealed so proper alignment of the intrinsic region with the PC cavity is critical.

\begin{figure}[!t]
\centering
\includegraphics[width=3.5in]{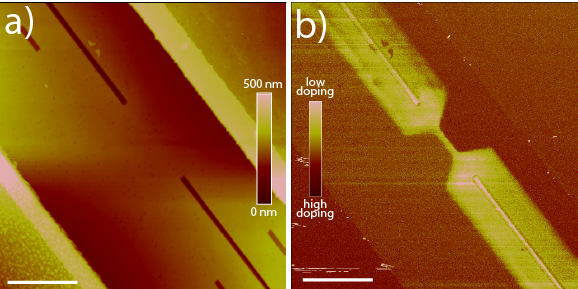}
\caption{(a) AFM topography image of the fabricated device without a photonic crystal. The scale bar is 5 $\mu$m. (b) SCM image of the region in (a). The p-side of the device is in the lower left corner and the n-side in the upper right. The trench is etched at the device centre, showing the precision of the alignment of the doping regions. `SCM data' is a combination of the phase and amplitude of capacitance data, where the strength of the signal is directly proportional to the intensity of doping in the local region under the tip \cite{williams}. The scale bar is 5 $\mu$m.}
\label{fig_sim}
\end{figure}

Our annealed QDs exhibit a blueshift by about 80 nm and a reduction in room temperature emission by a factor of ten (cryogenic emission strength is unchanged). Previous reports on annealed InAs quantum dots showed similar results with the likely conclusion that QD sizes undergo a redistribution and the QD core and wetting layer intermix \cite{malik}. Both effects decrease the QD carrier confinement potential, which has implications for room temperature operation, discussed in section IV.

We found that trenches surrounding the metal contact pads are vital in order to ensure current flows through only the cavity region. Due to light background doping of the top GaAs membrane, current can spread laterally through the entire membrane, bypassing the cavity and resulting in poor current-voltage characteristics. An example I-V plot of one of our former lateral diode devices fabricated without isolation trenches is shown in Figure 4. The quasi-linear current response is unrepresentative of a working diode and is more likely the result of the resistive membrane. The measured current level is in the mA range, which is orders of magnitude higher than what is expected and observed from properly functioning diodes with isolation trenches. As an alternative to the physical cuts imposed by the isolation trenches, future devices could potentially employ hydrogen implantation as a current aperture (as in VCSELs) in order to preserve the mechanical structure of the semiconductor membrane \cite{leeimplant}.   

\begin{figure}[!t]
\centering
\includegraphics[width=3.5in]{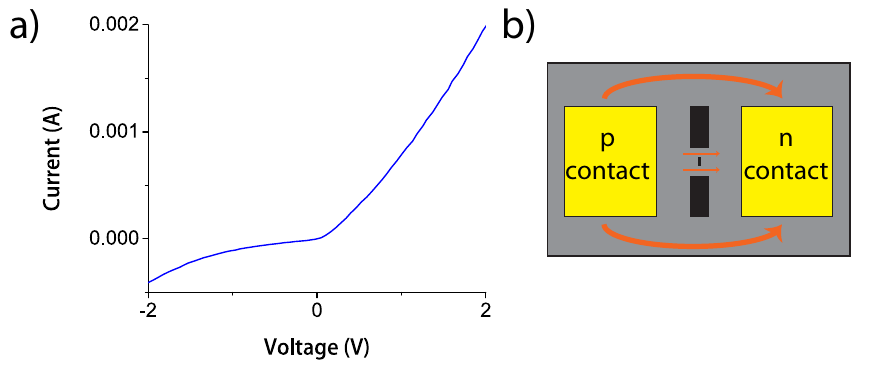}
\caption{(a) Typical I-V curve for a lateral junction fabricated without proper isolation trenches. The magnitude of the current is exceptionally high because of the large leakage pathway through the entire top GaAs membrane. The shape of the I-V curve does not resemble an ideal diode and is more characteristic of linear resistive behavior. (b) Diagram of the leakage current. Yellow squares are metal contacts and black rectangles are trenches. Most of the current is directed away from the cavity device region as seen by the orange arrows.}
\label{fig_sim}
\end{figure} 

Two-dimensional finite element Poisson simulations were carried out to predict the expected electrical behavior of our devices. Previous studies of electrical transport through InP photonic lattices have shown that current flow persists even in the presence of etched holes \cite{berrier}. The transport is modified by a geometric fill factor that accounts for the reduced membrane cross section area from the lattice holes and their respective depletion zones caused by Fermi level surface pinning. Though the exact nature of surface states requires a detailed analysis of trap energy levels, we approximate traps as acceptor type near mid-band and obtain good agreement between simulation and experiment. To find the appropriate trap density to use in simulation, we measure the material free carrier lifetime using a time-resolved photoluminescence setup and obtain a value of 6 ps (see section IV). This extremely short time of carrier relaxation is due to the non-radiative recombination from etched hole surfaces. From the non-radiative carrier lifetime, we back out a surface trap density of 5 $\times$ 10$^{13}$ cm$^{-2}$, and simulate our device to produce the current density and steady state electron and hole density plots in Figure 5. 

For a bias of 1.2 V, Figure 5(a) shows that the current flow is primarily through the L3 cavity region while minimal leakage current travels through the wide intrinsic mirrors. A unique current crowding effect is also visible around the cavity edges which approaches a large value of 10 kA/cm$^2$. The diode attains microamp level currents when biased near 1 V, and has a series resistance of roughly 1 k$\Omega$, dominated by the air-hole modified sheet resistance at this doping level (Figure 5(b)). Meanwhile the steady state e/h carrier densities saturate at around 10$^{16}$ cm$^{-3}$ at the cavity center (Figure 5(c,d)). The injection level is far lower than the nearby doping levels due to the high non-radiative recombination rate. While the above simulations were carried out at 300 K, we expect similar device electrical performance at lower temperatures with slightly lower currents and slightly higher carrier injection levels due to slower non-radiative recombination. 

To understand the effect of leakage current through the PC mirrors we also simulate devices having a wide 5 $\mu$m intrinsic region spacing and no taper profile for the doping. We see in Figure 5(b) that a small residual current does flow, but the magnitude is lower by a factor of ten. In Figure 6(a) an IR image of the EL from a wide intrinsic region device shows that indeed a small amount of emission is observed and is concentrated at the edge of the n-type doping region (this asymmetry in emission is due to the doping asymmetry and was confirmed via simulation). Even for devices with a normal tapered doping profile and sub-micron cavity intrinsic region, we observe finite leakage from the PC mirror intrinsic region sections as evidenced by the IR emission at the n-type region boundary (Figure 6(b)). The magnitude of this emission is much smaller compared to the cavity, however, as seen by the brightness of the signal. Comparing the I-V curves for these two devices we see that the wide intrinsic region device has a lower current by over an order of magnitude, as predicted by simulation. To avoid this small leakage in future devices, hydrogen implantation could again be used as a final current aperture, so long as the implantation damage is spaced at least a few microns away from the cavity. 

\begin{figure}[!t]
\centering
\includegraphics[width=3.5in]{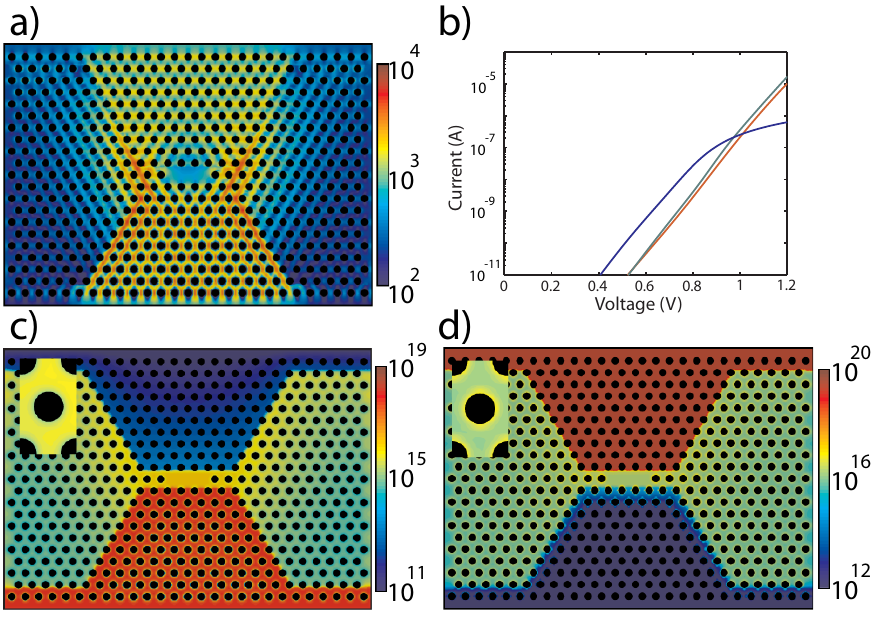}
\caption{(a) Simulated current density plot (in A/cm$^2$) of the L3 photonic crystal cavity design with a 400 nm intrinsic region width and a 5 $\mu$m outer mirror spacing. (b) Simulated I-V plots of the device with a 400 nm intrinsic region spacing and no holes (green), a 400 nm intrinsic region spacing with holes (red), and a device with only one large 5 $\mu$m wide intrinsic region with holes (blue). (c) Calculated eletron density map (in cm$^{-3}$) for a device with a 400 nm intrinsic region biased at 1.2 V. Inset shows a zoom-in of a hole region several periods away from the cavity. (d) Calculated hole density map (in cm$^{-3}$) for a device with a 400 nm intrinsic region biased at 1.2 V. Inset shows a zoom-in of a hole region several periods away from the cavity.}
\label{fig_sim}
\end{figure}

\begin{figure}[!t]
\centering
\includegraphics[width=3.5in]{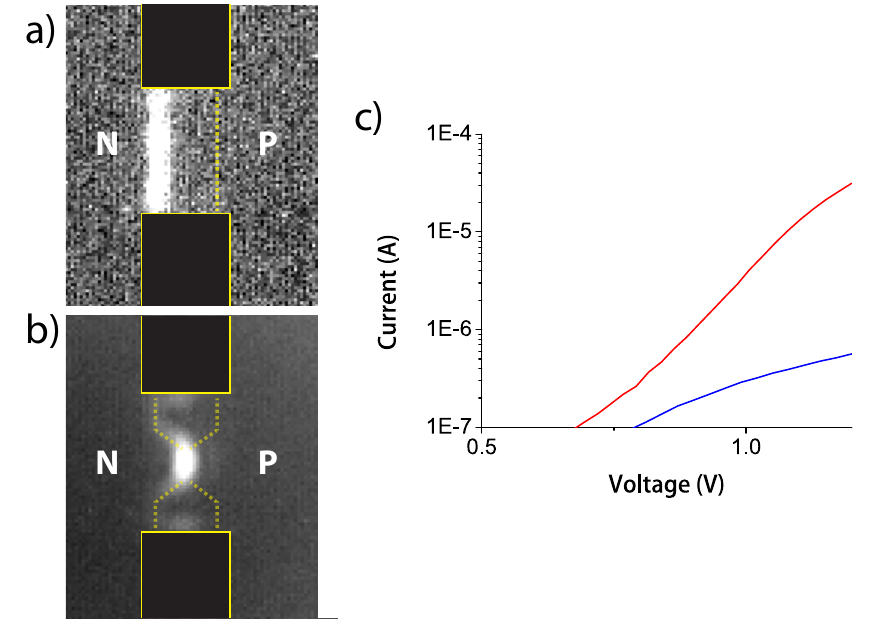}
\caption{(a) IR camera image of a room temperature LED (see section IV) with only a wide, 5 $\mu$m intrinsic region spacing. The doping is outlined in yellow dashed lines and trenches are shown by black rectangles. The emission, while weak, is still visible at the n-type region boundary. (b) IR camera image of a similar LED but with a tapered doping profile and with cavity intrinsic region separation of 400 nm. The EL is much brighter at the cavity compared to the n-type region boundary due to more efficient electrical injection. (c) Comparison of the I-V curves for the wide intrinsic region LED (blue trace) and the narrow intrinsic region LED (red trace).}
\label{fig_sim}
\end{figure}

\section{Ultralow threshold laser}

In this section we present results on our electrically pumped quantum dot PC laser \cite{ellislaser}. Our first attempt to produce such a laser using 900 nm InAs QDs was unsuccessful even at cryogenic temperatures \cite{ellisled}. For these shallow confinement QDs, the transparency carrier density is in excess of 10$^{17}$ cm$^{-3}$. With previous activated dopant levels in the 10$^{18}$ range, the injected carrier concentration in the cavity is much lower than that needed for inversion and gain as per the discussion in section II. Therefore it is not surprising that stimulated emission was not observed for our former devices.     

For our lasing structure, we use a GaAs membrane with three layers of high density (300 dots/$\mu$m$^2$) InAs QDs with peak emission strength near 1300 nm. These deep confinement QDs have a much lower transparency carrier density of 5 $\times$ 10$^{14}$ cm$^{-3}$ and should acheive inversion with our electrical scheme. The parameters of the cavity are chosen so that the fundamental cavity mode is at a wavelength of 1174 nm at low temperature, within the ground state emission of the quantum dots. Figure 7 shows optical and electron microscope pictures of a fully fabricated device. Figure 8(b) is an I-V plot of our laser device. From the curve, we observe a dominant exponential current (the red line) corresponding to current flowing through the cavity along with a leakage component at low biases. We found that for this set of devices the leakage was due to an incompletely removed AlGaAs sacrificial layer and subsequent devices did not have this problem (see section IV).  

\begin{figure}[!t]
\centering
\includegraphics[width=3.5in]{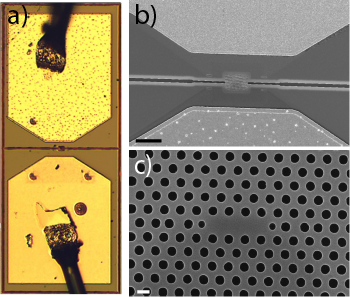}
\caption{(a) Optical microscope image of a complete device showing large metal contact pads along with the connecting wire bonds. The PC diode is located in the center between the contact pads. (b) SEM image of a fabricated laser diode device. The p-side of the device appears on the top of the image, and the n-side on the bottom. The scale bar is 10 $\mu$m. (c) SEM image of the photonic crystal cavity (zoom-in of the central region of (b)). The scale bar is 300 nm.}
\label{fig_sim}
\end{figure}

Figure 8(a) shows the current in-light out properties of our laser at 50K and 150K. We observe a clear lasing threshold for temperatures below 150 K. We find that the threshold of our laser is 181 nA at 50 K and 287 nA at 150 K (corresponding to 208 nW and 296 nW of consumed power). To the best of our knowledge, this is the lowest threshold ever demonstrated in any electrically pumped laser. It is three orders of magnitude better than the 260 $\mu$A threshold of previous quantum well PC cavity lasers and more than an order of magnitude better than the thresholds of metal-clad lasers and micropost lasers \cite{park1, hill, reitz}. The low thresholds demonstrated in these lasers are a result of the optimized lateral current pathway, where charge can be efficiently delivered to the cavity center. From the collected power, we estimate the total power radiated by the laser to be on the order of a few nW well above threshold. The inset in Figure 8(a) shows the experimental far-field radiation pattern of the cavity above threshold, showing a clear speckle pattern.

\begin{figure}[!t]
\centering
\includegraphics[width=3.5in]{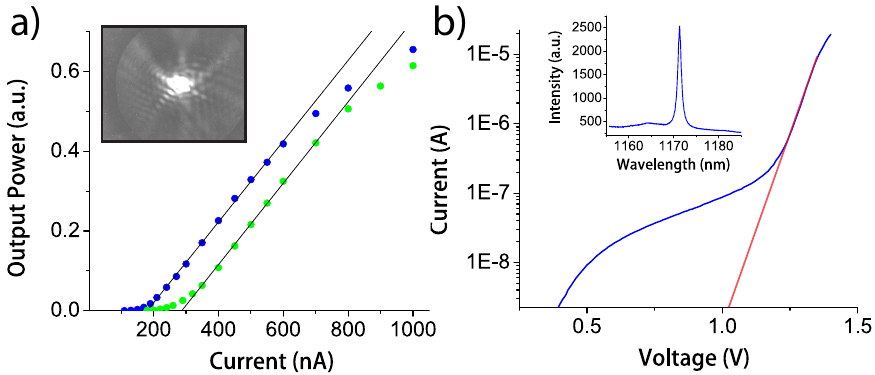}
\caption{(a) Experimental current-light characteristics for the laser at 50 K (blue points) and 150 K (green points). The black lines are linear fits to the above threshold output power of the lasers, which are used to find the thresholds. The inset is a far field radiation pattern of the laser at a current of 3 $\mu$A. (b) I-V plot for the laser device. The initial hump is leakage current not flowing through the cavity. The red line shows the current that is flowing through the device. Inset shows the lasing spectrum above threshold.}
\label{fig_sim}
\end{figure}

At low voltages before the diode has fully turned on we observe leakage current bypassing the cavity through residual AlGaAs material that was not fully etched away. Therefore, if the AlGaAs was fully removed for this device, the threshold could be significantly lower. To find the potential threshold reduction, we fit the current voltage characteristics to an ideal diode equation to determine the fraction of current flowing through the cavity (as shown in Figure 8(b)). The laser threshold after this leakage current correction is only 70 nA.

\section{Ultrafast single-mode LED}

Room temperature operation of our nanocavity light source reveals interesting physical properties that can be exploited for ultrafast modulation \cite{shambatled}. Though our devices do not lase, they exhibit effectively single-mode LED behavior with QD emission coupled to cavity resonances and can be directly modulated at very high speeds. Devices were fabricated as described previously with the only difference being a longer undercut step to eliminate substrate leakage current. A fabricated diode is seen in Figure 9(a) with a corresponding I-V curve and output emission spectrum for a forward bias of 10 $\mu$A (Figure 9(b)). The current is slightly greater than at cryogenic temperatures because of increased non-radiative recombination current at room temperature. Additionally substrate leakage current is no longer observed because of the optimized undercut step. Bright and clearly defined cavity modes peak well above the background QD emission with the fundamental mode centered at 1260 nm and having a Q-factor of 1600.  

\begin{figure}[!t]
\centering
\includegraphics[width=3.5in]{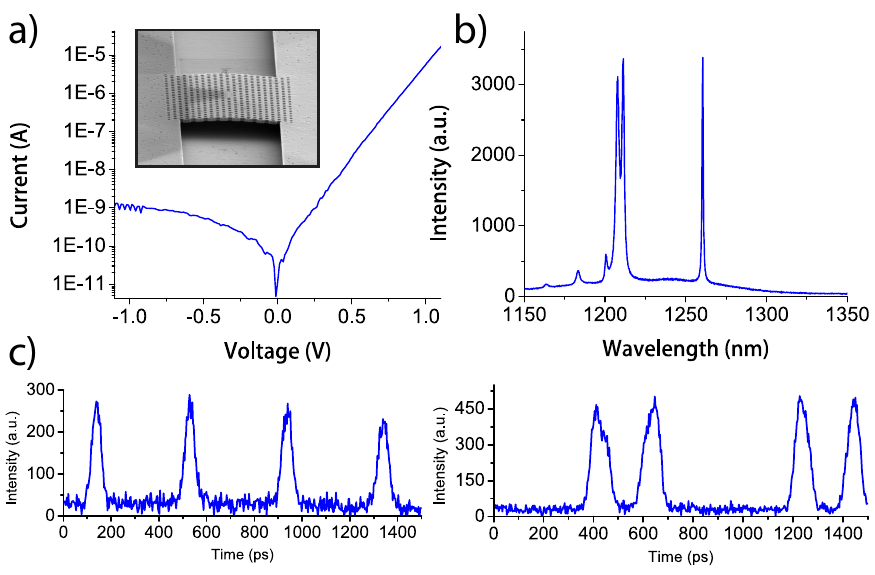}
\caption{(a) Diode current-voltage plot measured for a typical LED device. Leakage current is minimized for this set of room temperature devices. The inset shows an SEM image of a fabricated device. (b) Spectrum of the cavity for a forward bias current of 10 $\mu$A. The fundamental mode for this device is at 1260 nm. (c) Direct modulation results of the room temperature single-mode LED for two different pattern sequences showing ultrafast operation.}
\label{fig_sim}
\end{figure}

Time resolved lifetime measurements using a Ti:Sapphire laser as an optical pump were studied to examine the recombination rates in our system. At room temperature, the nominal QD lifetime is 100-300 ps due to bulk non-radiative recombination \cite{shambatled}. The quantum dots in our active LED devices undergo a rapid thermal annealing step at 850~$^{\circ}$C which causes the QD ensemble emission to blueshift by 80 nm and decrease in intensity by ten-fold. Likely, the QD core semiconductor intermixes with the wetting layer surroundings and there is a narrowing of the overall QD ensemble sizes during the anneal \cite{malik}. These effects would induce a shallower confining potential and hence promote a more rapid reemission of carriers from the quantum dots. For our annealed PC devices, we measure a QD lifetime of only 10 ps at 1100 nm, indicating rapid escape of carriers from our modified QDs and subsequent fast non-radiative recombination from etched surfaces \cite{shambatled}. 

We perform direct electrical modulation studies of our LEDs next to determine the time-resolved response of our devices. Figure 9(c) shows the single-mode output for two different bit patterns fed to the diode through a pulse pattern generator. The diode replicates the voltage pulse patterns very well and has pulse widths of 100 ps, limited by our pattern generator. Fast electrical data is thereby mapped onto the single mode carrier of the nanocavity LED, which can be used as the light source for optical interconnect transmission.

Our device is over an order of magnitude faster with three orders of magnitude lower power consumption (here only 2.5 $\mu$W) compared to previously shown directly modulated photonic crystal single-mode LEDs at cryogenic temperatures \cite{francardi}. The power consumption for the 10 GHz non-return-to-zero speed diode in Figure 9 is only 2.5 $\mu$W, indicating an average energy per bit of only 0.25 fJ \cite{shambatled}. Power output for this particular LED is quite low in the 10s to 100s of pW owing to the fast non-radiative recombination, but is still within the range of detection for advanced photodiode technology \cite{assefa, hayden}. 

\section{Additional applications of lateral junction active devices}

\subsection{Nanobeam photonic crystal LED}

One-dimensional (1D) nanobeam cavities have emerged recently as a competing technology to two-dimensional photonic crystal membrane technology. As light is confined in these cavities by distributed Bragg reflection in only one dimension, they can have a smaller footprint and higher quality factors than their two-dimensional counterparts. Recently researchers have been able to demonstrate high quality factors in 1D nanobeam cavities in low-index materials such as silicon dioxide and silicon nitride \cite{gong1, gong2}. They have also been used to demonstrate low threshold lasers, optomechanical crystals, and chemical sensors \cite{zhang, eichenfeld, wang}. We show that in spite of their narrow cross sections, nanobeam cavities can be efficiently electrically pumped by a lateral p-i-n junction \cite{shambatnanobeam}. 

Figure 10(a) shows SEM images of a fabricated, laterally doped photonic crystal nanobeam. Figure 10(b) is an electroluminescence spectrum of a representative nanobeam device when biased to 5 $\mu$A. The electroluminescence IR image corresponding to Figure 10(b) is shown in Figure 10(c). Together, these results demonstrate that the current is efficiently directed by the lateral p-i-n junction to the cavity region. We find that as the injection current of the cavity is increased a small amount of differential gain is observed, indicating lasing in this cavity design is possible. These results represent a promising advance towards practical active nanobeam device architectures.

\begin{figure}[!t]
\centering
\includegraphics[width=3.5in]{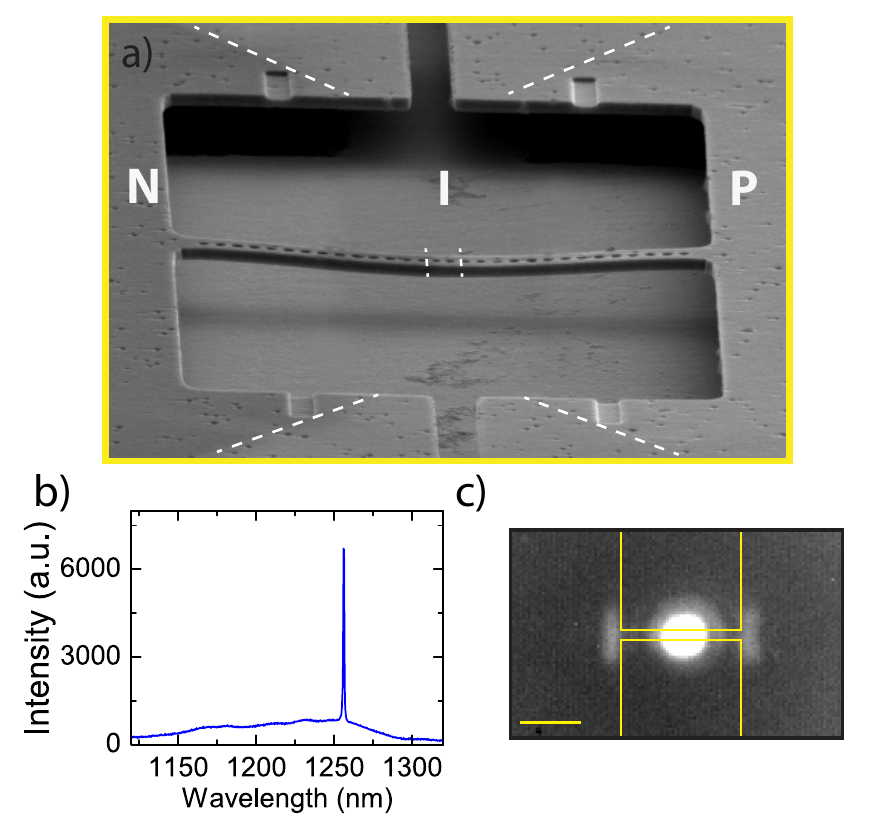}
\caption{(a) Tilted SEM image of a nanobeam structure. The n-type doping is seen as darker grey and the doping regions are outlined in white dashed lines. The beam is deflected down by a small amount likely due to strain from the GaAs/AlGaAs interface. (b) EL spectrum for a nanobeam device at a forward bias of 5 $\mu$A. The cavity fundamental mode is the sharp peak at 1255 nm and the background QD emission is the broad spectrum below. The Q-factor is found to be 2900. (c) Nanobeam cavity emission taken with an IR camera. An outline of the cavity is seen by the yellow lines and the scale bar is 5 $\mu$m. The cavity emission is bright at the center as expected and there is slight EL scattered out at the nanobeam edges.}
\end{figure}

\subsection{Fiber-taper coupled electro-optic modulator}

We further test the usefulness of our lateral p-i-n structure by demonstrating an electro-optic modulator based out of a cavity coupled to a fiber taper waveguide \cite{shambatmod}. GaAs has a much stronger free carrier dispersion than does silicon and has fast carrier non-recombination at surfaces. Together with the small mode volume of the cavity of  0.7($\lambda$/n)$^3$, over an order of magnitude smaller than typical microring resonators \cite{lipson}, these advantages enable ultra-low power operation at potentially very high speeds. The electrically contacted devices were fabricated as discussed above with cavities centered near 1500 nm in a passive GaAs membrane. Fiber tapers were fabricated as before having a 1 $\mu$m diameter and only a few dB of loss \cite{shambattaper}. 

Modulation is achieved by first aligning the fiber taper to the cavity (Figure 11(a)) such that the fundamental mode resonance of the PC cavity appears as a dip in transmission when light is sent through the fiber. Figure 11(b) shows 100 MHz operation using a p-i-n detector with an RF amplifier circuit, proving high speed operation by free carrier dispersion. The ultimate switching speed for an injection based diode is given by the carrier lifetime (6 ps), suggesting an ultimate speed of up to 100 GHz here. We find that our switching energy is only 0.6 fJ/bit \cite{shambatmod}, again confirming the low power advantage of our nanocavity.

\begin{figure}[!t]
\centering
\includegraphics[width=3.5in]{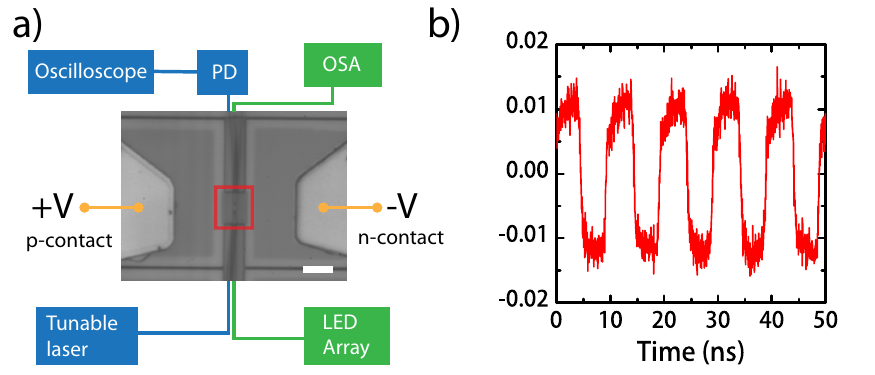}
\caption{(a) Schematic for taper-based modulator experiment with underlying optical image of a fiber taper aligned to a cavity. PD is photodiode and OSA is optical spectrum analyzer. The cavity region is outlined in the red box. The scale bar is 10 $\mu$m. (b) 100 MHz modulation of our device, detected using a p-i-n photodetector with an RF amplifier circuit.}
\end{figure}

\subsection{Fiber-coupled photodetector}

The fiber-taper coupled to an electrically contacted photonic crystal can be used to demonstrate a cavity-enhanced two-photon photodetector. Cavities have been previously used to enhance the responsivity of photodetectors in both the linear and two-photon regimes \cite{kishino,tanabe2}. Conceptually, a cavity can provide a longer effective path length (for linear absorption) as well as an increased local light intensity (for two-photon absorption (TPA)). Ordinarily, GaAs will have zero linear absorption near 1.5 $\mu$m due to its larger bandgap; however its nonlinear TPA coefficient is 10 cm GW$^{-1}$, which is over ten times greater than the TPA coefficient of silicon \cite{nozakitpa}. Therefore, two-photon absorption and subsequent photodetection may be possible in GaAs without the need for extremely high Q-factor cavities.  
   
We perform measurements on our previously fabricated lateral junctions at 1500 nm. As before, the fiber taper is aligned and coupled to the L3 cavity (Figure 12(a)). The transmission coupling wavelength is noted and a tunable laser matched to the cavity resonance is fed into one taper end. Current-voltage traces are then taken for various input laser powers. We see in Figure 12(b) that the device indeed functions as a two-photon detector with photocurrent increasing proportionally to the input laser power. By incorporating the taper loss as well as the power coupling ratio into the cavity, we estimate the responsivity to be around 10$^{-3}$ A/W. Photodetection occurs even with our low Q factors of 1000-2000, and is seen only when the pump laser is within the cavity bandwidth, confirming resonant enhancement. The absolute responsivity could be further improved by using cavities with higher Q-factors. 

\begin{figure}[!t]
\centering
\includegraphics[width=3.5in]{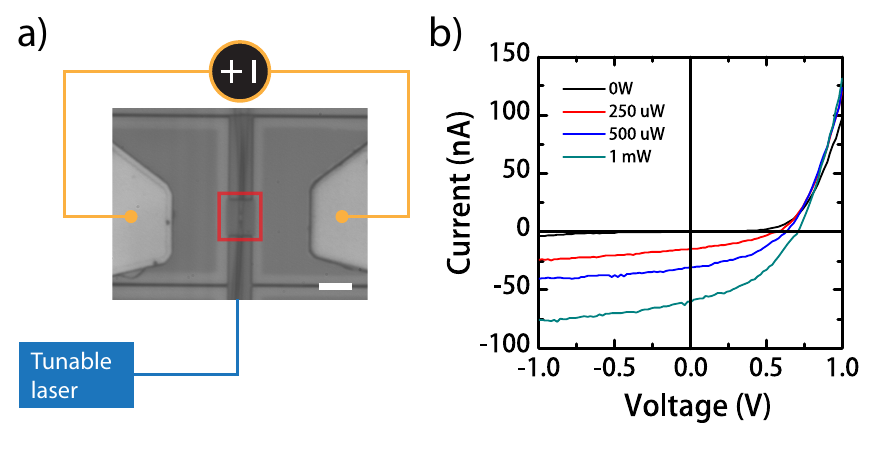}
\caption{(a) Schematic for the fiber-coupled photodetector. The taper is aligned as previously and the photocurrent is monitored as a laser is fed into one end of the fiber. The cavity region is outlined in the red box. The scale bar is 10 $\mu$m. (b) Photocurrent IV plots for several input laser powers.}
\end{figure}

\subsection{Interconnected lateral junction devices}

We conclude our results by discussing multiply interconnected lateral junction devices. Practical devices in future on-chip networks will need to be multiplexed at high density, requiring a minimum footprint for all components: lasers, modulators, and detectors. For high data rates to be possible, wavelength division multiplexing is necessary, requiring many independent light channels. A recent study illustrated this concept by multiplexing several independent ring resonator modulators at different wavelengths to one coupled waveguide \cite{lipsonwdm}. In our devices, we are able to reproduce this behavior with multiple cavities embedded in the same PC lattice.

For our structures, we fabricate one large PC lattice with three uncoupled cavities spaced by six or more lattice periods in passive GaAs material (Figure 13(a-c)). The number of linked cavities can be increased further but for the present demonstration we use three. The local hole radius and lattice constant can be tuned so that the cavity modes have identical or different wavelengths. An example where different wavelengths are desirable would be for having multiplexed sources or modulators for different channels. Identical wavelengths might be useful for matching a source with a modulator, a source with a detector, etc. The doping layout is such that large intrinsic regions separate adjacent cavities (Figure 13(c)). This is done so that an applied bias to one set of contacts does not produce a crosstalk current flow in a neighboring cavity. Finally, trenches are fabricated to isolate each metal contact pad to eliminate crosstalk (Figure 13(a)).

To couple all three cavities we again use a fiber taper waveguide that is appropriately placed on all three cavities. When positioned properly, we obtain clear transmission signals for each cavity (Figure 13(d)). In this example, the three cavity modes were fabricated to have unique wavelengths near 1500 nm for modulation studies. As a forward bias is applied to a single pair of contact pads, the corresponding cavity mode is seen to shift in accordance to the modulation properties discussed above. Meanwhile the other two cavity modes are unaffected by the applied bias. This is repeated with the other two cavity modes with the same result. Therefore, independent electrical control without crosstalk can be achieved in a compact platform. Extensions of this design can easily be made with any number of active components for dense optoelectronic integration.     

\begin{figure}[!t]
\centering
\includegraphics[width=3.5in]{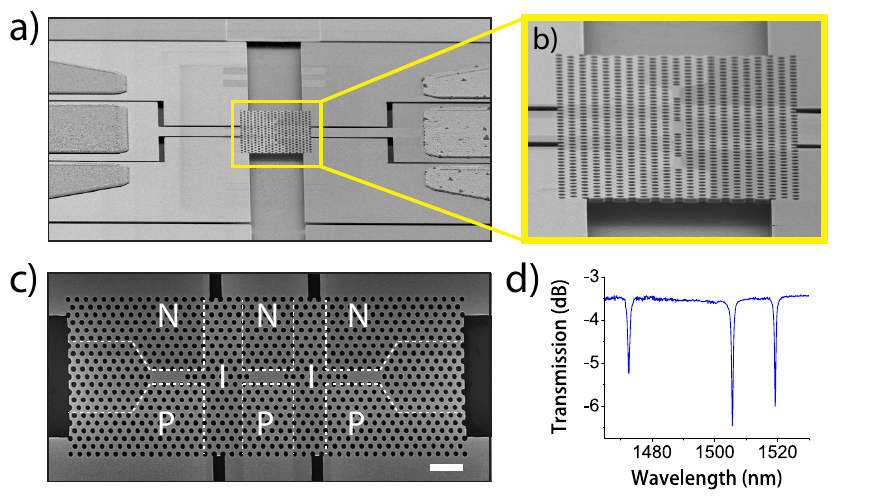}
\caption{(a) Tilted SEM picture of a fabricated triple cavity device. The independent contact pads are seen along with the isolation trenches. (b) Zoom-in SEM of (a). The doping profile is partially visible. (c) Top-down SEM of a device with doping regions labeled and delineated by dashed lines. The scale bar is 2 $\mu$m. (d) Transmission spectrum for a fiber taper coupled to a set of three cavities. The fundamental mode was designed to be at a unique wavelength for each cavity. Independent tuning of each mode is possible.}
\end{figure}

\section{Discussion and outlook}

We have demonstrated a novel platform for the electrical control of active photonic crystal nanocavity devices. In particular, we have achieved efficient electrical injection into 2D membranes of GaAs containing quantum dots for laser and LED applications. Our laser operating at 50K has a threshold of only 181 nA, the lowest of any electrically driven laser ever demonstrated. The lasers operate up to 150K before transitioning to LEDs due to limited QD gain and high cavity loss. 

The principal reason why we do not observe lasing at room temperature is that the quality factor of our cavities is too low, and hence the cavity loss is too great compared to the supplied QD gain. Previous reports on InAs quantum dot lasers showed that room temperature lasing in this material system is quite challenging due to the low maximum gain provided by high density QDs, quoted as up to 5 cm$^{-1}$ per layer \cite{nomura, srinivasan}. In order to realize lasing with such low gain media, the cavity loss must be exceptionally small with Q-factors in excess of 10$^4$. Our devices, on the other hand, have Q factors ranging from 1000-2000, limited by several fabrication complexities.

The first is that the doping regions partially overlap with the cavity mode field, resulting in free carrier absorption. This loss is unavoidable since the p and n regions must be closely spaced in order to ensure proper carrier injection (see Section II). The second limitation to the Q-factor we believe is due to some surface roughening of the GaAs membrane caused by the repeated deposition and etching of the nitride masks. An additional challenge to room temperature lasing is that the InAs QDs are strongly affected by the rapid thermal anneal at 850 $^{\circ}$C. Their emission strength goes down by a factor of ten at room temperature, likely due to reduced carrier confinement in the dots (see Section IV). While it may be possible to optimize the fabrication process to simultaneously increase the Q-factor and maintain the original QD gain, it will not be trivial. 

Alternative to the GaAs platform, indium phosphide with embedded quantum wells might prove to be a better material system for achieving room temperature lasing. QWs have significantly more gain than do QDs and room temperature lasing in InP is routinely observed even for low Q cavities \cite{painter3, loncar2, altug2}. Quantum wells have even been shown to have increased photoluminescence intensity after rapid thermal annealing \cite{lee}. The main challenge with transferring this platform to InP is the relatively high transparancy carrier density needed for QW inversion (typically around 10$^{18}$ cm$^{-3}$ \cite{altug2}). As mentioned in section II, achieving high carrier densities at the cavity center is limited by activated doping levels and non-radiative recombination. Implantation in InP should result in similar activation levels as for GaAs, though the Si (n dopant) would have a higher activation compared to Be (p dopant) \cite{rao}. At the same time, non-radiative recombination for InP has been shown to be over an order of magnitude slower due to the positioning of surface trap defects within the bandgap \cite{nolte, holzman}. Our preliminary Poisson simulations show that it should be possible to attain inversion in QWs and lasing at room temperature in properly designed InP lateral junctions. If this can be done, practical low-threshold lasers with high output powers and fast modulation rates could become a true reality for on-chip photonics.

The uniquely modified properties of our quantum dots at room temperature have allowed us to demonstrate a 10 GHz single-mode LED with sub-fJ/bit energy consumption. While a traditional laser is typically the device of choice for an optical source, our single-mode LED has all the look and feel of a laser and can be used for practical communications. The cavity allows light to be efficiently channeled into a well-defined mode that can be further extracted with a coupled waveguide. The narrow spectral linewidth also allows for wavelength division multiplexing in densely integrated systems. Low output power remains an issue in these types of sources relying on non-radiative recombination alone. A fast, single-mode LED that is also very efficient by having a large Purcell enhancement would require sophisticated engineering of the material system and careful optimization of cavity parameters \cite{wu}. We point out, however, that even though our LEDs are low power, optical interconnect links could be designed around such transmitters by using sensitive detector technology.  

Finally, we have utilized our lateral p-i-n junction to demonstrate LEDs in 1D photonic crystal nanobeams and fiber-coupled modulators and detectors. The former work is an encouraging result in that electrical injection can be surprisingly efficient for such small structures. Nanobeams have conducting cross-sectional areas of just a few hundred square nanometers, yet high current densities and carrier injection levels similar to those of 2D membranes are possible. 

Our electro-optic modulator and two-photon photodetector present alternative functions for the lateral p-i-n structure not related to light emission. Here we show that a cavity can be controlled through free carrier dispersion and in the process modulate light output through a coupled fiber taper. Cavity-enhanced TPA and photocurrent generation open up another avenue for electrical PC cavity control. Going forward, integrated components such as the triple cavity for wavelength division multiplexing will be important for full system applications.    

In summary, we have developed a technique for both efficient electrical injection and electrical control of photonic crystal cavities. Design flexibility is inherent in our platform and numerous parameters can be easily changed to provide a custom need. The extra degree of freedom provided through electrical control could lead to new physics and studies on the interaction between light and matter, as well as in optomechanics. 

\section*{Acknowledgment}

Gary Shambat and Bryan Ellis were supported by the Stanford Graduate Fellowship. Gary Shambat is also supported by the NSF GRFP. The authors acknowledge the financial support of the Interconnect Focus Center, one of the six research centers funded under the Focus Center Research Program, a Semiconductor Research Corporation program. We also acknowledge the AFOSR MURI for Complex and Robust On-chip Nanophotonics (Dr. Gernot Pomrenke), grant number FA9550-09-1-0704, and the Director, Office of Science, Office of Basic Energy Sciences, Materials Sciences and Engineering Division, of the US Department of Energy under Contract No. DE-AC02-05CH11231. Work was performed in part at the Stanford Nanofabrication Facility of NNIN supported by the National Science Foundation. We also acknowledge Kelley Rivoire for assisting in SEM image acquisition.

\ifCLASSOPTIONcaptionsoff
  \newpage
\fi


\begin{thebibliography}{1}

\bibitem{jelena1}
J.~Vu\v{c}kovi\'{c}, M.~Lon\v{c}ar, H.~Mabuchi, and A.~Scherer, ``Optimization of Q-factor in photonic crystal microcavities," \emph{IEEE J. Quantum. Elec.}, vol. 38, pp. 850-856, 2002.

\bibitem{akahane1}
Y.~Akahane, T.~Asano, B.S.~Song, and S.~Noda, ``High-Q photonic nanocavity in a two-dimensional photonic crystal," \emph{Nature}, vol. 425, pp. 944-947, 2003.

\bibitem{noda1}
B.S.~Song, S.~Noda, T.~Asano, and Y.~Akahane, ``Ultra-high-Q photonic double-heterostructure nanocavity," \emph{Nat. Mater.}, vol. 4, pp. 207-210, 2005.

\bibitem{noda2}
Y.~Tanaka, J.~Upham, T.~Nagashima, T.~Sugiya, T.~Asano, and S.~Noda, ``Dynamic control of the Q factor in a photonic crystal nanocavity," \emph{Nat. Mater.}, vol. 6, pp. 862-865, 2007.

\bibitem{painter1}
J.~Chan, T.~P.~M.~Alegre, A.~H.~Safavi-Naeini, J.~T.~Hill, A.~Krausse, S.~Groblacher, M.~Aspelmeyer, and O.~Painter, ``Laser cooling of a nanomechanical oscillator into its quantum ground state," \emph{Nature}, vol. 478, pp. 89-92, 2011.

\bibitem{jelena2}
D.~Englund, A.~Faraon, I.~Fushman, N.~Stoltz, P.~Petroff, and J.~Vu\v{c}kovi\'{c}, ``Controlling cavity reflectivity with a single quantum dot," \emph{Nature}, vol. 450, pp. 857-861, 2007.

\bibitem{fan1}
J.~S.~Foresi, P.~R.~Villeneuve, J.~Ferrera, E.~R.~Thoen, G.~Steinmeyer, S.~Fan, J.~D.~Joannopoulos, L.~C.~Kimmerling, H.~I.~Smith, and E.~P.~Ippen, ``Photonic-bandgap microcavities in optical waveguides," \emph{Nature}, vol. 390, pp. 143-145, 1997.

\bibitem{painter2}
O.~Painter, J.~Vu\v{c}kovi\'{c}, and A.~Scherer, ``Defect modes of a two-dimensional photonic crystal in an optically thin dielectric slab," \emph{J. Opt. Soc. Am. B}, vol. 16, pp. 275-285, 1999.

\bibitem{painter3}
O.~Painter, R.~K.~Lee, A.~Scherer, A.~Yariv, J.~D.~O'Brien, P.~D.~Dapkus, I.~Kim, ``Two-dimensional photonic band-gap defect mode laser," \emph{Science}, vol. 284, pp. 1819-1821, 1999.

\bibitem{noda3}
S.~Noda, ``Photonic crystal lasers - ultimate nanolasers and broad area coherent lasers," \emph{J. Opt. Soc. Am. B}, vol. B 27, pp. B1-B8, 2010.

\bibitem{yamamoto}
G.~Bjork and Y.~Yamamoto, ``Analysis of semiconductor microcavity lasers using rate equations," \emph{IEEE J. Quantum Elec.}, vol. 27, pp. 2386-2396, 1991.

\bibitem{strauf}
S.~Strauf, K.~Hennessy, M.~T.Rakher, Y.~S.~Choi, A.~Badolato, L.C.~Andreani, E.~L.~Hu, P.~M.~Petroff, and D.~Bouwmeester, ``Self-tuned quantum dot gain in photonic crystal lasers," \emph{Phys. Rev. Lett.}, vol. 96, pp. 127404, 2006.

\bibitem{matsuo}
S.~Matsuo, A.~Shinya, T.~Kakitsuka, K.~Nozaki, T.~Segawa, T.~Sato, Y.~Kawaguchi, and M.~Notomi, ``High-speed ultracompact buried heterostructure photonic-crystal laser with 13 fJ of energy consumed per bit transmitted," \emph{Nat. Photonics}, vol. 4, pp. 648-654, 2010.

\bibitem{altug}
H.~Altug, D.~Englund, and J.~Vu\v{c}kovi\'{c}, ``Ultra-fast photonic crystal nanolasers," \emph{Nat. Physics}, vol. 2, pp. 484-488, 2006.

\bibitem{nomura}
M.~Nomura, S.~Iwamoto, K.~Watanabe, N.~Kumagai, Y.~Nakata, S.~Ishida, and Y.~Arakawa, ``Room temperature continuous-wave lasing in photonic crystal nanocavity," \emph{Opt. Express}, vol. 14, pp. 6308-6315, 2006.

\bibitem{nozaki}
K.~Nozaki, H.~Watanabe, and T.~Baba, ``Photonic crystal nanolaser monolithically integrated with passive waveguide for effective light extraction," \emph{Appl. Phys. Lett.}, vol. 92, pp. 021108, 2008.

\bibitem{park1}
H.~G.~Park, S.~H.~Kim, S.~H.~Kwon, Y.~G.~Ju, J.~K.~Yang, J.~H.~Baek, S.~B.~Kim, and Y.~H.~Lee, ``Electrically driven single-cell photonic crystal laser," \emph{Science}, vol. 305, pp. 1444-1447, 2005.

\bibitem{park2}
H.~G.~Park, S.~H.~Kim, M.~K.~Seo, Y.~G.~Ju, S.~B.~Kim, and Y.~H.~Lee, ``Characteristics of electrically driven two-dimensional photonic crystal laser," \emph{IEEE J. Quantum Elec.}, vol. 41, pp. 1131-1141, 2005.

\bibitem{macdougal}
M.~H.~Macdougal, P.~D.~Dapkus, V.~Pudikov, H.~Zhao, and G.~M.~Yang, ``Ultralow threshold current vertical-cavity surface-emitting lasers with AlAs oxide-GaAs distributed Bragg reflectors," \emph{IEEE Photon. Technol. Lett.}, vol. 7, pp. 229-231, 1995.

\bibitem{francardi}
M.~Francardi, L.~Balet, A.~Gerardino, N.~Chauvin, D.~Bitauld, L.~H.~Li, B.~Alloing, and A.~Fiore, ``Enhanced spontaneous emission in a photonic-crystal light-emitting diode," \emph{Appl. Phys. Lett.}, vol. 93, pp. 143102, 2008.

\bibitem{chakravarty}
S.~Chakravarty, P.~Bhattacharya, J.~Topol'an\v{c}ik, and Z.~Wu, ``Electrically injected quantum dot photonic crystal microcavity light emitters and microcavity arrays," \emph{J. Phys. D: Appl. Phys.}, vol. 40, pp. 2683-2690, 2007.

\bibitem{ellisled}
B.~Ellis, T.~Sarmiento, M.~A.~Mayer, B.~Zhang, J.~Harris, E.~E.~Haller, and J.~Vu\v{c}kovi\'{c}, ``Electrically pumped photonic crystal nanocavity light sources using a laterally doped p-i-n junction," \emph{Appl. Phys. Lett.}, vol. 96, pp. 181103, 2010.

\bibitem{ellislaser}
B.~Ellis, M.~A.~Mayer, G.~Shambat, T.~Sarmiento, James~S.~Harris, E.~E.~Haller, and J.~Vu\v{c}kovi\'{c}, ``Ultralow-threshold electrically pumped quantum dot photonic-crystal nanocavity laser," \emph{Nat. Photonics}, vol. 5, pp. 297-300, 2011.

\bibitem{loncar1}
I.~W.~Frank, P.~B.~Deotare, M.~W.~McCutcheon, and M.~Lon\v{c}ar, ``Programmable photonic crystal nanobeam cavities," \emph{Opt. Express}, vol. 18, pp. 8705-8712, 2010.

\bibitem{andrei1}
A.~Faraon, A.~Majumdar, H.~Kim, P.~Petroff, and J.~Vu\v{c}kovi\'{c}, ``Fast electrical control of a quantum dot strongly coupled to a photonic crystal cavity," \emph{Phys. Rev. Lett.}, vol. 104, pp. 047402, 2010.

\bibitem{lipson}
Q.~Xu, B.~Schmidt, S.~Pradhan, and M.~Lipson, ``Micrometre-scale silicon electro-optic modulator," \emph{Nature}, vol. 435, pp. 325-327, 2005.

\bibitem{tanabe1}
T.~Tanabe, K.~Nishiguchi, E.~Kuramochi, and M.~Notomi, ``Low power and fast electro-optic silicon modulator with lateral p-i-n embedded photonic crystal nanocavity," \emph{Opt. Express}, vol. 17, pp. 22505-22503, 2009.

\bibitem{tanabe2}
T.~Tanabe, H.~Sumikura, H.~Taniyama, A.~Shinya, and M.~Notomi, ``All-silicon sub-Gb/s telecom detector with low dark current and high quantum efficiency on chip," \emph{Appl. Phys. Lett.}, vol. 96, pp. 101103, 2010.

\bibitem{watanabe}
H. Watanabe and T. Baba, ``Active/passive-integrated photonic crystal slab micro-laser," \emph{Elec. Lett.}, vol. 42, pp. 695-696, 2006.

\bibitem{tager}
A.~Tager, R.~Gaska, I.~Avrutzky, M.~Fay, H.~Chik, A.~SpringThorpe, S.~Eicher, J.~M.~Xu, and M.~Shur, ``Ion-implanted GaAs-InGaAs lateral current injection laser," \emph{IEEE J. Sel. Top. Quant.}, vol. 5, pp. 664-672, 1999.

\bibitem{long}
C.~M.~Long, A.~V.~Giannopoulos, and K.~D.~Choquette, ``Modified spontaneous emission from laterally injected photonic crystal emitter," \emph{Elec. Lett.}, vol. 45, pp. 227-228, 2009.

\bibitem{englund}
D.~Englund, H.~Altug, and J.~Vu\v{c}kovi\'{c}, ``Low-threshold surface-passivated photonic crystal nanocavity laser," \emph{Appl. Phys. Lett.}, vol. 91, 071124, 2007.

\bibitem{rao}
M.~V.~Rao, ``Ion implantation in III-V compound semiconductors," \emph{Nucl. Instrum. Meth. B.}, vol. 79, 645-647, 1993.

\bibitem{williams}
C.~C.~Williams, ``Two-dimensional dopant profiling by scanning capacitance microscopy," \emph{Ann. Rev. Mater. Sci.}, vol. 29, 471-504, 1999.

\bibitem{shambatnanobeam}
G.~Shambat, B.~Ellis, M.~A.~Mayer, J.~Petykiewicz, T.~Sarmiento, James~S.~Harris, E.~E.~Haller, and J.~Vu\v{c}kovi\'{c}, ``Nanobeam photonic crystal cavity light-emitting diodes," \emph{Appl. Phys. Lett.}, vol. 99, 071105, 2011.

\bibitem{malik}
S.~Malik, C.~Roberts, R.~Murray, and M.~Pate, ``Tuning self-assembled InAs quantum dots by rapid thermal annealing," \emph{Appl. Phys. Lett.}, vol. 71, 1987-1989, 1997.

\bibitem{leeimplant}
Y.~H.~Lee, B.~Tell, K.~Brown-Goebeler, J.~L.~Jewell, and J.~V.~Hove, ``Top-surface-emitting GaAs four-quantum-well lasers emitting at 0.85 $\mu$m," \emph{Elec. Lett.}, vol. 26, 710-711, 1990.

\bibitem{berrier}
A.~Berrier, M.~Mulot, G.~Malm, M.~Ostling, and S.~Anand, ``Carrier transport through a dry-etched InP-based two-dimensional photonic crystal," \emph{J. Appl. Phys.}, vol. 101, 123101, 2007.

\bibitem{hill}
M.~T.~Hill, Y.~S.~Oei, B.~Smalbrugge, Y.~Zhu, T.~de~Vries, P.~J.~van~Veldhoven, F.~W.~M.~van~Otten, T.~J.~Eijkemans, J.~P.~Turkiewicz, H.~de~Waardt, E.~J.~Geluk, S.~H.~Kwon, Y.~H.~Lee, R.~Notzel, and M.~K.~Smith, ``Lasing in metallic-coated nanocavities," \emph{Nat. Photonics}, vol. 1, pp. 589-594, 2007.

\bibitem{reitz}
S.~Reitzenstein, T.~Heindel, C.~Kistner, A.~Rahimi-Iman, C.~Schneider, S.~Hofling, and A.~Forchel, ``Low threshold electrically pumped quantum dot-micropillar lasers," \emph{Appl. Phys. Lett.}, vol. 93, 061104, 2008.

\bibitem{shambatled}
G.~Shambat, B.~Ellis, A.~Majumdar, J.~Petykiewicz, M.~A.~Mayer, T.~Sarmiento, James~S.~Harris, E.~E.~Haller, and J.~Vu\v{c}kovi\'{c}, ``Ultrafast direct modulation of a single-mode photonic crystal nanocavity light-emitting diode," \emph{Nat. Comm.}, vol. 2, 539, 2011.

\bibitem{assefa}
S.~Assefa, F.~Xia, and Y.~A.~Vlasov, ``Reinventing germanium avalanche photodetector for nanophotonic on-chip optical interconnects," \emph{Nature}, vol. 464, 80-85, 2010.

\bibitem{hayden}
O.~Hayden, R.~Agarwal, and C.~M.~Lieber, ``Nanoscale avalanche photodiodes for highly sensitive and spatially resolved photon detection," \emph{Nat. Mater.}, vol. 5, 352-356, 2006.

\bibitem{gong1}
Y.~Gong and J.~Vu\v{c}kovi\'{c}, ``Photonic crystal cavities in silicon dioxide," \emph{Appl. Phys. Lett.}, vol. 96, 031107, 2010.

\bibitem{gong2}
Y.~Gong, M.~Makarova, S.~Yerci, R.~Li, M.~J.~Steven, B.~Baek, S.~W.~Nam, L.~Dal~Negro, and J.~Vu\v{c}kovi\'{c}, ``Observation of transparency of erbium-doped silicon nitride in photonic crystal nanobeam cavities," \emph{Opt. Express}, vol. 18, 12176-12184, 2010.

\bibitem{zhang}
Y.~Zhang, M.~Khan, Y.~Huang, J.~H.~Ryou, P.~B.~Deotare, R.~Dupuis, and M. Lon\v{c}ar, ``Photonic crystal nanobeam lasers," \emph{Appl. Phys. Lett.}, vol. 97, 051104, 2010.

\bibitem{eichenfeld}
M.~Eichenfield, J.~Chan, R.~Camacho, K.~J.~Vahala, and O.~Painter, ``Optomechanical crystals," \emph{Nature}, vol. 462, 78-82, 2009.

\bibitem{wang}
B.~Wang, M.~A.~Dundar, R.~Notzel, F.~Karouta, S.~He, and R.~W.~van~der~Heijden, ``Photonic crystal slot nanobeam slow light waveguides for refractive index sensing," \emph{Appl. Phys. Lett.}, vol. 97, 151105, 2010.

\bibitem{shambatmod}
G.~Shambat, B.~Ellis, M.~A.~Mayer, A.~Majumdar, E.~E.~Haller, and J.~Vu\v{c}kovi\'{c}, ``Ultra-low power fiber-coupled gallium arsenide photonic crystal cavity electro-optic modulator," \emph{Opt. Express}, vol. 19, 7530-7536, 2011.

\bibitem{shambattaper}
G.~Shambat, Y.~Gong, J.~Lu, S.~Yerci, R.~Li, L.~Dal~Negro, and J.~Vu\v{c}kovi\'{c}, ``Coupled fiber taper extraction of 1.53 $\mu$m photoluminescence from erbium doped silicon nitride photonic crystal cavities," \emph{Opt. Express}, vol. 18, 5964-5973, 2010.

\bibitem{kishino}
K.~Kishino, M.~S.~Unlu, J.~I.~Chyi, J.~Reed, L.~Arsenault, and H.~Morkoc, ``Resonant cavity-enhanced (RCE) photodetectors," \emph{IEEE J. Quantum Elec.}, vol. 27, 2025-2034, 1991.

\bibitem{nozakitpa}
K.~Nozaki, T.~Tanabe, A.~Shinya, S.~Matsuo, T.~Sato, H.~Taniyama, and M.~Notomi, ``Sub-femtojoule all-optical switching using a photonic-crystal nanocavity," \emph{Nat. Photonics}, vol. 4, 477-483, 2010.

\bibitem{lipsonwdm}
S.~Manipatruni, L.~Chen, and M.~Lipson, ``Ultra high bandwidth WDM using silicon microring modulators," \emph{Opt. Express}, vol. 18, 16858-16867, 2010.

\bibitem{srinivasan}
K.~Srinivasan, M.~Borselli, T.~J.~Johnson, P.~E.~Barclay, and O.~Painter, ``Optical loss and lasing characteristics of high-quality-factor AlGaAs microdisk resonators with embedded quantum dots," \emph{Appl. Phys. Lett.}, vol. 86, 151106, 2005.

\bibitem{loncar2}
M.~Lon\v{c}ar, T.~Yoshie, A.~Schere, P.~Gogna, and Y.~Qiu, ``Low threshold photonic crystal laser," \emph{Appl. Phys. Lett.}, vol. 81, pp. 2680, 2002.

\bibitem{altug2}
H.~Altug and J.~Vu\v{c}kovi\'{c}, ``Photonic crystal nanocavity array laser," \emph{Opt. Express}, vol. 13, 8819-8828, 2005.

\bibitem{lee}
C.~Y.~Lee, M.~C.~Wu, Y.~D.~Tian, W.~H.~Wang, W.~J.~Ho, and T.~T.~Shi, ``Effects of rapid thermal annealing on InAsP/InP strained multiquantum well laser diodes grown by metal organic chemical vapour deposition," \emph{Elec. Lett.}, vol. 36, 1026-1028, 2000.

\bibitem{nolte}
D.~D.~Nolte, ``Surface recombination, free-carrier saturation, and dangling bonds in InP and GaAs," \emph{Solid State Electron.}, vol. 33, 295-298, 1990.

\bibitem{holzman}
J.~F.~Holzman, P.~Strasser, R.~Wuest, F.~Robin, D.~Erni, and H.~Jackel, ``Ultrafast carrier dynamics in InP photonic crystals," \emph{Nanotechnology}, vol. 16, 949-952, 2005.

\bibitem{wu}
E.~K.~Lau, A.~Lakhani, R.~S.~Tucker, and M.~C.~Wu, ``Enhanced modulation bandwidth of nanocavity light emitting devices," \emph{Opt. Express}, vol. 17, 7790-7999, 2009.

\end{thebibliography}
\end{document}